\documentclass[aps,prl,twocolumn,superscriptaddress]{revtex4-1}
\usepackage{graphicx}
\usepackage{anysize}
\usepackage{lineno}
\usepackage[amssymb]{SIunits}
\usepackage{lipsum}
\usepackage{amsmath}
\usepackage{amssymb}
\usepackage[toc,page]{appendix}
\usepackage{SIunits}
\usepackage{soul}
\usepackage[dvipsnames]{xcolor}
\usepackage{bm}

\newcommand \be {\begin{equation}}
\newcommand \ee {\end{equation}}
\newcommand \bea {\begin{eqnarray}}
\newcommand \eea {\end{eqnarray}}

\renewcommand{\vec}[1]{\bm{{#1}}}
\renewcommand{\tensor}[1]{\bm{{#1}}}

\newcommand{\B}{\mathcal{B}}
\newcommand{\XB}{\vec{X}_{\B}}

\newcommand{\XT}{\vec{X}_{t}}

\usepackage{ulem}
\definecolor{thibautcolor}{HTML}{FF517B}

\definecolor{checkcolor}{HTML}{FF003F}

\begin{document}

%\preprint{}

\title{Universal Scaling Laws for a Generic Swimmer Model}



\author{Bruno Vent\'ejou}
\email[]{bruno.ventejou@univ-grenoble-alpes.fr}
\affiliation{Univ. Grenoble Alpes, CNRS, LIPhy, 38000 Grenoble, France}
\author{Thibaut M\'etivet}
\email[]{thibaut.metivet@inria.fr}
\affiliation{Univ. Grenoble Alpes, Inria, CNRS, Grenoble INP, LJK, 38000 Grenoble, France}
\author{Aur\'elie Dupont}
\affiliation{Univ. Grenoble Alpes, CNRS, LIPhy, 38000 Grenoble, France}
\author{Philippe Peyla}
\email[]{philippe.peyla@univ-grenoble-alpes.fr}
\affiliation{Univ. Grenoble Alpes, CNRS, LIPhy, 38000 Grenoble, France}

% \date{\today}

\begin{abstract}
We have developed a minimal model of a swimmer without body deformation based on force and torque dipoles which allows accurate 3D Navier-Stokes calculations. 
Our model can reproduce swimmer propulsion for a large range of Reynolds numbers, and generate wake vortices in the inertial regime, reminiscent of the flow generated by the flapping tails of real fish.
We performed a numerical exploration of the model from low to high Reynolds numbers and obtained
universal laws using scaling arguments. 
We collected data from a wide variety of micro-organisms, thereby extending the experimental data presented in (M. Gazzola et al., Nature Physics 10, 758, 2014).  Our theoretical scaling laws compare very well with experimental data across the different regimes, from Stokes to turbulent flows. 
We believe that this model, due to its relatively simple design, will be very useful for obtaining numerical simulations of collective effects within fish schools composed of hundreds of individuals.

\end{abstract}

\maketitle

\textbf{Introduction}
The wide variety of means employed by living creatures to move in aquatic environments is fascinating \cite{Childress1981}. Motion generally involves a complex interplay between the deformation of the body and the surrounding fluid. From the smallest organisms, like bacteria, to colossal blue whales \cite{Berg2004,Powar2022,garcia2011random,Smits2019,Gray1936,Wolfgang1999,Dabiri2006, Muller2001}, the differences in length scale $L$, velocity $v$ and mode of locomotion are so vast that the elaboration of a universal model to describe swimming across these scales might seem impossible.

The importance of inertia with respect to viscous dissipation is quantified by the Reynolds number, which expresses the ratio of stress due to inertia to stress due to viscosity: $Re=\rho v L / \eta$, where $\rho$ and $\eta$ represent fluid density and viscosity, respectively.
The Reynolds numbers associated with aquatic living species span several decades, typically ranging from $Re^{\rm fish} \sim 10^3 - 10^6$ for fish to $Re^{\rm micro-org} \lesssim 10^{-3}$ for micro-organisms like spermatozoa. Consequently, the drag force exerted by the fluid on the swimmer is largely dependent on the species considered, originating from either fully viscous dissipation at the small scale, or turbulent inertia at the large scale.

As a result, each swimming model \cite{Taylor1951,Spagnolie2010,Lauga2020,Lighthill1960,Liu1999,Kern2006} tends to offer a tailored approach specific to the flow regime considered and the corresponding body deformation. 
Most models focus on periodic body deformation, coupled with the surrounding fluid, and resolve the full swimming cycle. 
This provides specific approaches that address swimming at small scales, where viscous forces dominate (e.g. for micro-organisms~\cite{LaugaPowers2009}), differently from the macroscopic strategies of large fish or mammals, which can leverage the inertia of the surrounding fluid to break time reversibility~\cite{Purcell1977}. 
The highly diverse physical origins of particular swimming patterns represent an obstacle to the exploration of a more comprehensive and universal viewpoint.
%\ppnote{Note that recently, a model based on point like swimmers has been developed in an invisicd 2D potential fluid \cite{Filella2018}. }

In this letter, we propose a different approach. Deformation kinematics, such as undulations, oscillations and pulsations, are ignored, and locomotion is described using force and torque dipoles applied by a solid body of finite size $L$ on a fluid. While a similar description has already been used for micro-swimmers at low $Re$~\cite{Hernandez2005,Mehandia2008,Jibuti2014}, to the best of our knowledge, it has never yet been employed for high $Re$, when inertia starts to dominate viscous forces. In this work, we study the motion described by our model over $8$ decades of Reynolds numbers ($10^{-5} \lesssim Re \lesssim 10^4$), theoretically and numerically, comparing our results with experimental data.  
Although the direct effect on swimming velocity of the specific deformation of the swimmer’s body is acknowledged, particularly as it can reduce the drag force~\cite{Gray1936,Li2021}, we would like to emphasize that we are not attempting to provide a detailed and precise analysis of a particular mode of locomotion, but rather a more universal description in terms of the forces applied to the fluid. While our swimmer model is minimal, the motion of the surrounding fluid is accurately captured using the full numerical resolution of the 3D Navier-Stokes equation, and our approach encompasses the different swimming regimes of a wide variety of aquatic species. Our model can also remarkably reproduce the characteristic wake vortices observed behind fish due to the flapping of their tails \cite{Drucker2002}. 
%We show that this wake is more and more dissipated while going to small Reynolds numbers. In addition, we show that our model swimmer also possesses steering properties and is able to perform controllable trajectories.

%One of the force is applied on the body  itself while the other one is applied in the fluid behind the body such mimiking a flagellum, a tail or any swimming oragan. 

%In model $2$, both forces and torques are periodic functions representing the periodic stroke of the swimmer at pulsation $\omega$. 
%In model $1$, we consider the time averaging on one period $2\pi / \omega$ of the force, the averaging of the torque being zero. 

Our approach furthermore exhibits universal scaling laws which link the swimming Reynolds number $Re$ to a new dimensionless group, the thrust number defined below.
%: $Th=\rho f_0 L^3/\eta^2$ \ppnote{where $F=f_0 L^3$ is the force exerted by the body of the swimmer on the fluid (Fig. \ref{schema}) and $f_0$ the force density. This number compares both inertial forces and applied forces to viscous forces (see below)}. 
We identify three different regimes: the Stokes regime ($Re<1$), a laminar regime ($1<Re<10^3$) and a turbulent regime ($Re > 10^3-10^4$), and calculate the theoretical exponents of the scaling laws in the three regimes using simple scaling analyses (independently of the space dimension). These compare very well with the numerical simulations produced by our generic swimmer model. We also validate our results with experimental data presented in~\cite{Gazzola2014} for the laminar and turbulent regimes, and further extend this validation with data collected on micro-swimmers for the Stokes regime.
%, and Concerning the laminar and the turbulent regimes, we show that our force model is validated by experimental results based on kinematics data on a thousand of aquatic organisms and also by numerical simulations \cite{Gazzola2014}. For the Stokes flow, we have collected data for microswimmers that also validate our results.

\begin{figure}
	\centerline{\includegraphics[width=\linewidth]{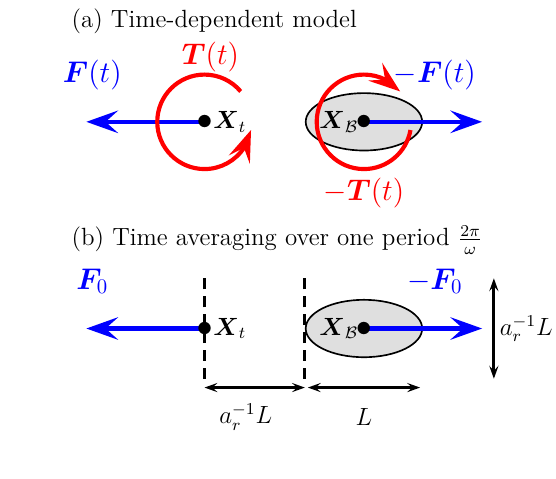}}
        \caption{The swimmer is an ellipsoidal rigid body of length $L$ and width $L/a_r$, with $a_r$ the aspect ratio. a)~Time-dependent dipoles of forces $\{-\vec{F}, \vec{F}\}$ and torques $\{-\vec{T}, \vec{T}\}$ are applied at the center-of-mass of the body $\XB$ and the ``phantom tail'' $\XT$, with $\bm{F}(t)=(\pi/2) \bm{F}_0 \lvert \cos (\omega t) \rvert$  and $\bm{T}(t)=\bm{T}_0 \cos (\omega t)$. The position of $\XT$ can be controlled to steer the swimmer, but is kept such that $\| \XB - \XT \| = L/2 + L/a_r$. b)~Static model obtained by averaging the time dependent model.}
	\label{schema}
\end{figure}

\begin{figure}
    \centering
    \includegraphics[width=\linewidth]{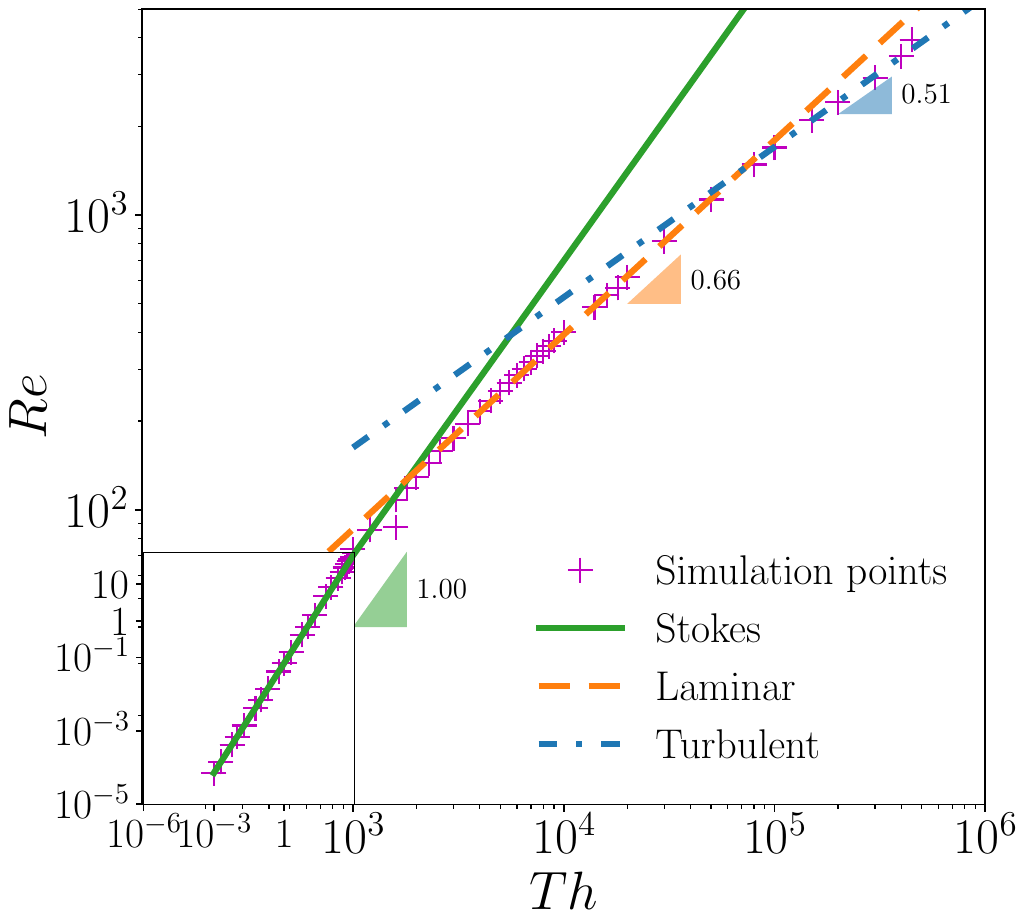}
    \caption{$Re$ as a function of $Th$ from 3D numerical simulations (crosses), obtained by solving the full Navier-Stokes equations with our swimmer model. We clearly obtain three regimes: $Re \sim Th^{1.0}$ for $Re\lesssim20$; $Re \sim Th^{0.66}$ for $20\lesssim Re \lesssim1000$ and $Re \sim Th^{0.51}$ for $Re \gtrsim 1000$. The three lines correspond to fitting curves and give the numerical scaling exponents. Note that the crossovers between the different regimes depend on the geometry of the swimmer; this curve corresponds to $a_r = 4$.}
    \label{fig:RevsTh}
\end{figure}

\paragraph{\bf Swimmer model.} The model uses a time-dependent force dipole combined with a torque dipole (Fig.~\ref{schema}(a)), both attached to a rigid body $\B$ of ellipsoidal shape. An autonomous swimming body creates its own motion, therefore the total sum of forces and torques must cancel out, due to the third law of Newton of the fluid-body system.~\cite{Feynman2006}. 
The model developed by \citet{Filella2018} presents some conceptual similarities. It represents each fish as a point-like active particle bearing a dipole in a potential 2D inviscid fluid, which allows consideration of the hydrodynamic interactions between fish in the far-field limit. However, since our aim is to explore a wide spectrum of Reynolds numbers (from viscous to inertial regimes), we solved the full incompressible Navier-Stokes equations in 3D and 2D (see SM \cite{ventejou24SM}).
%(in 2D and 3D) in the presence of a solid body carrying dipoles of forces and torques. In addition, the finite size of the swimmer allows us to obtain a good hydrodynamic signature of the fish \adnote{not reduced to \sout{not only in} the far field \sout{approximation}}. For example, in inertial regimes, our model is able to generate a vortex alley in the swimmer's wake similar to the one observed behind a fish. }

%The swimmer is a rigid body of elliptical shape with a long and short axis of size $L$ and $L/2$ respectively. 
The swimmer exerts on the fluid a force dipole  $\{\bm{-F},\bm{F}\}$ like that generally used for a micro-swimmer at a low $Re$ \cite{Hernandez2005,Mehandia2008,Jibuti2014} (see Fig.~\ref{schema}). We used a pusher-like model that reproduces the force distribution of a fish at high Reynolds numbers. This approach can easily be extended to a more detailed model by using more complex force distributions (\textit{e.g.} \cite{Mehandia2008}). As shown in Fig.~\ref{schema}(a), the force dipole is composed of one force applied in the fluid at the rear of the body $\XT$, mimicking a swimming organ, and an opposite force exerted inside the body at the center-of-mass $\XB$. The force is time-dependent with pulsation $\omega$ and pusher-like: $\bm{F}(t)=(\pi/2)\bm{F_0}  \lvert \cos \,\omega t \rvert$ with $\mathbf{F}_0 \cdot (\XB - \XT) > 0$.
The absolute value in the expression of $\bm{F}(t)$ enforces the pusher nature of the swimmer.  
We also consider a torque dipole: $\bm{T}(t)=\bm{T_0} \cos \,\omega t$, collocated with the force dipole. The torque at the back represents the stroke of the swimming organ and causes the vortex street \cite{Drucker2002} in the fish’s wake at high $Re$. An opposite torque is applied in the body (Fig.~\ref{schema}(a)), and represents the counter-reaction of the rest of the body. For practical reasons in the scaling analyses below, we also introduce force and torque densities $f_0 \equiv F_0/L^3$ and $\tau_0 \equiv T_0 / L^3$ respectively. Averaging these dipoles over one period of time ($2\pi/\omega$) results in a simple static force dipole $\{\bm{-F_0},\bm{F_0}\}$ (Fig.~\ref{schema}(b)), while the average torque cancels out. This time averaging over one beating period is very similar to models of pushers and pullers beating at low $Re$ \cite{Hernandez2005,Mehandia2008,Jibuti2014}.
%\thibaut{Note that the model proposed by \cite{Filella2018} corresponds to the limit of infinite $Re$ and only addresses far-field hydrodynamics interactions of point-like particles.}
%We study both $2D$ and $3D$ models. We show that the emerging scaling laws are independent of the dimension of space (see SM for the $2D$ case). 

%Thus respecting that the total internal efforts (forces and torques) exerted by the swimmer on the fluid cancel out for an isolated and autonomous body (i.e. in the absence of any external field of forces such as gravity for example). 

\textbf{Scaling laws}. The hydrodynamic nature of our model allows for simple scaling arguments, inspired by \citet{Gazzola2014} for inertial flows but translated to a more generic framework and extended to the non-inertial Stokes regime. In the following, we present $3D$ arguments, but they remain valid in $2D$ (see SM \cite{ventejou24SM}). We also consider that all the lengths scale as $L$. 

The body of the swimmer is submitted to different dominant drag forces depending on the Reynolds number.
To describe this effect across all swimming regimes, we introduce the thrust number $Th$ as the ratio between the applied force density $f_0$ multiplied by inertial forces $\rho \lvert D \mathbf{v}/Dt \rvert \sim \rho v^2 L^{-1}$, and the square of viscous forces $\lvert \eta \Delta \mathbf{v} \rvert ^2 \sim (\eta v L^{-2})^2$, which gives
\begin{equation}
    Th \equiv \frac{\rho f_0 L^3}{\eta^2}.
    \label{DefTh}
\end{equation}
The thrust number appears naturally at all scales of Reynolds numbers, as shown below. It contains the force term at the origin of the motion, which is characterized by the Reynolds number. It therefore provides a convenient method for evaluating velocity as a function of force. Let us consider classic scaling arguments:

\begin{itemize}
    \item At high Reynolds numbers, the boundary layer around the body is turbulent and pressure drag dominates. The corresponding force scales as $f_0 L^3 \sim \rho v^2 L^2$. Since $v \sim \eta Re/ (\rho L)$,  we obtain $Re \sim (\rho f_0 L^3/\eta^2)^{1/2}=Th^{1/2}$.
    \item For small but finite Reynolds numbers, the regime is laminar and the viscous force in the boundary layer dominates: $f_0 L^3 \sim (\eta v/\delta) L^2$ where $\delta$ is the thickness of the boundary layer, which obeys the Blasius law \cite{landau2013fluid} $\delta \sim L \,Re^{-1/2}$. This finally leads to $Re \sim Th^{2/3}$.
    \item At low Reynolds numbers, the Stokes drag force dominates and the force applied on the body compensates the drag:  $f_0 L^3 \sim \eta v L$. This gives $Re \sim \rho f_0 L^3/\eta^2=Th$.
\end{itemize}
From the Stokes to the turbulent regime, we observe that the exponent $\alpha$ of the scaling $Re\sim Th^\alpha$ is always below one and decreases. It suggests a diminishing swimming performance as the Reynolds number of the swimmer increases.
To confirm these three successive regimes, we present below numerical simulations with our swimmer model, exploring a large range of values for the thrust and Reynolds numbers.

\paragraph{\bf Numerical simulations.} We performed direct numerical simulations of the incompressible Navier-Stokes equations in the presence of our model swimmer, a rigid ellipsoid body with the force and torque dipoles attached. The corresponding fluid momentum balance equation writes:
\begin{equation}
    \begin{aligned}
        \rho \frac{D\bm{v}}{Dt} - \nabla \cdot \left( 2\eta \tensor{E}(\vec{v}) \right)+ \nabla p
        = \vec{\mathcal{F}} + \vec{\mathcal{T}}
        %=\bm{f}(\bm{r},t)+\frac{1}{2}\bm{\nabla} \times \bm{\tau}(\bm{r},t),
    \end{aligned}
    \label{eq:NavierStokesEq}
\end{equation}
where $\vec{v}$ and $p$ are respectively the velocity and pressure fields, $D/Dt$ denotes the material derivative $D\vec{v}/Dt \equiv \partial \vec{v}/\partial t+(\bm{v\cdot\nabla})\bm{v}$, $\tensor{E}(\vec{v}) \equiv (\bm{\nabla v}+\bm{\nabla v}^t)/2$ is the strain-rate tensor, $\rho$ and $\eta$ denote the density and viscosity fields, and $\vec{\mathcal{F}}(t)$ and $\vec{\mathcal{T}}(t)$ respectively represent the -- time-varying -- force and torque dipoles attached to the swimmer, as introduced above.
The rigid body $\mathcal{B}$ of the swimmer is accounted for with a fictitious domain penalty method inspired by~\citep{janela2005penalty}; in practice, this can be implemented simply with a spatially-variable viscosity~\citep{Tanaka2000}: $\eta = \eta_f + (\eta_b - \eta_f) \, H_{\mathcal{B}}$ where $H_{\mathcal{B}}$ is the indicator -- or Heaviside -- function of $\mathcal{B}$. In practice, a viscosity ratio $\eta_b/\eta_f=10^3-10^6$ is applied between the fluid and the swimmer body to ensure that the rigid motion constraint $\tensor{E}(\vec{v}) = 0$ is satisfied within $\mathcal{B}$. Density $\rho$ is defined as constant inside and outside $\mathcal{B}$, thus making the swimmer neutrally buoyant.
Note that this fictitious domain penalty method allows the swimmer to be treated directly as part of the Navier-Stokes equations through the viscosity field, thereby avoiding the need to deal with moving boundary conditions and potential remeshing issues at the body interface in the discrete setting. The incrompressible Navier-Stokes equations are solved numerically using an implicit $\mathcal{P}2-\mathcal{P}1$ finite element method~\citep{metivet2018high} implemented in the parallel FEEL++ library~\cite{Prudhomme2012}. 
The position and orientation of the swimmer are updated at each time-step using a first order Euler scheme with the translational and rotational velocities computed from the fluid velocity field in $\mathcal{B}$.
A comprehensive derivation of the numerical model and technical details are provided in SM \cite{ventejou24SM}.

This numerical framework is used to explore a large range of Reynolds numbers ($10^{-5}<Re<10^{4}$) and Thrust numbers ($10^{-3}<Th<10^6)$, in order to evaluate the dependence of $Re$ as a function of $Th$ while varying each of the model’s different parameters ($L$, $\eta_f$, $f_0$, $\omega$ and $\tau_0$) separately~(see SM~\cite{ventejou24SM}). 
As shown in Fig.~\ref{fig:RevsTh}, the numerical simulations are in perfect agreement with the scaling laws presented above, displaying the Stokes regime in the range $10^{-5} \lesssim Re \lesssim 10^2$, the laminar regime for $10^2 \lesssim Re \lesssim 10^3$ and the turbulent regime for $10^3 \lesssim Re$, 
\begin{figure}[h]
	\centering
    \includegraphics[width=0.95\linewidth]{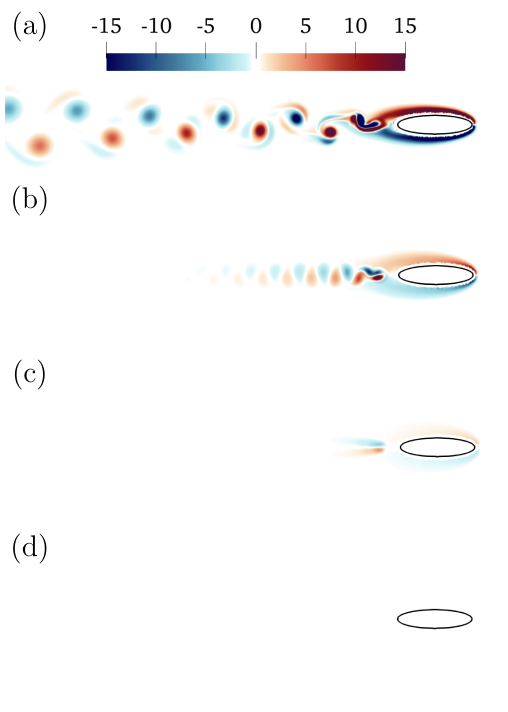}
    \caption{2D Vorticity field for different $Re$. The simulations were performed with $\omega=2\pi$, $L=32$ and $\| \vec{F_0} \| = \| \vec{T_0} \|$. (a): $\lVert \bm{T}_0 \rVert=4000$ and $Re=960$. (b): $\lVert \bm{T}_0 \rVert=600$ and $Re=300$.  (c): $\lVert \bm{T}_0 \rVert=100$ and $Re=50$.  (d): $\lVert \bm{T}_0 \rVert=0.25$ and $Re=0.1$.
 % $Re=530$ $T_0=6000$ $F_0=6000$ (b): $Re=170$ $T_0=800$ $F_0=800$ (c): $Re=10^{-2}$ $T_0=1$ $F_0=0.2$. \bvnote{A completer} \adnote{on peut mettre le Re dans la figure?}\bvnote{DISCUTER LE PREFACTEUR}}
 % $Re=530$ (b): $Re=170$ (c): $Re=10^{-2}$\bvnote{CHECK NUMERICAL VALUES USED} 
 }
	\label{vorticityField}
\end{figure}
with fitting exponents that match the predictions. 
Note that the $Re$ ranges of each regime depend on the aspect ratio of the swimmer $a_r$, which is kept constant. We also found that $\omega$ and $\tau_0$ do not play any role in the $Re(Th)$ dependency, which confirms that all the important parameters are embedded in the $Th$ number.
% We also plot $Re(Th)$ for the static dipole (Fig.~\ref{schema}-b) which is obtained by averaging the force and torque dipoles on a period $2\pi/\omega$.  
% We re-obtain the same scaling as before (Fig.~\ref{fig:RevsTh})\thibaut{WHICH ONE? I ONLY SEE THE SCALING AVERAGED DIPOLES}, thus confirming that $\omega$ and $\tau_0$ do not play any role in the scaling at constant $f_0$. Note that the crossovers between the different regimes vary as a function of the aspect ratio of the swimmer ($\alpha$ parameter in Fig.~\ref{schema}) \adnote{(c'est curieux, on a l'impression que tu dis deux fois la même chose. qu'est-ce qui a été fait avec le modèle complet?}.

%We will see below that a relation exists between $f_0$ and $\omega$ which allows us to recover some already published experimental results \cite{Gazzola2014} on the scaling laws between the Reynolds number $Re$ and the so-called swimming number $Sw=\rho \omega L^2/\eta$.  This validates our model that we extent to the Stokes regime

\textbf{Wake and vortices}. Although torque plays no role in the scaling of $Re$ as a function of $Th$, it is essential to reproduce the wake at the rear of the swimmer in the inertial regime. The torque dipole can be used to account for flagellum, body undulation or tail beating to generate a reverse von Karman vortex street, as observed in the wake of a fish at high $Re$ number \cite{Kern2006, Ko2023}. 

% {\it i.e.} an alley of vortices \adnote{(on dit deux fois la même chose)}. 

% Each vortex represents the beating of a tail in one direction and in the other, generating an alley of successive vortices of opposite sign (Fig.~\ref{vorticityField}). 
% Figure~\ref{vorticityField} clearly shows the vortex alley created behind the swimmer in inertial regimes.
Figures~\ref{vorticityField}(a,b) illustrate the typical wakes obtained with $(T_0,F_0)=(4000,4000),(600,600)$ in 2D simulations. 
\begin{figure}[ht]
    \centering
    \includegraphics[width=\linewidth]{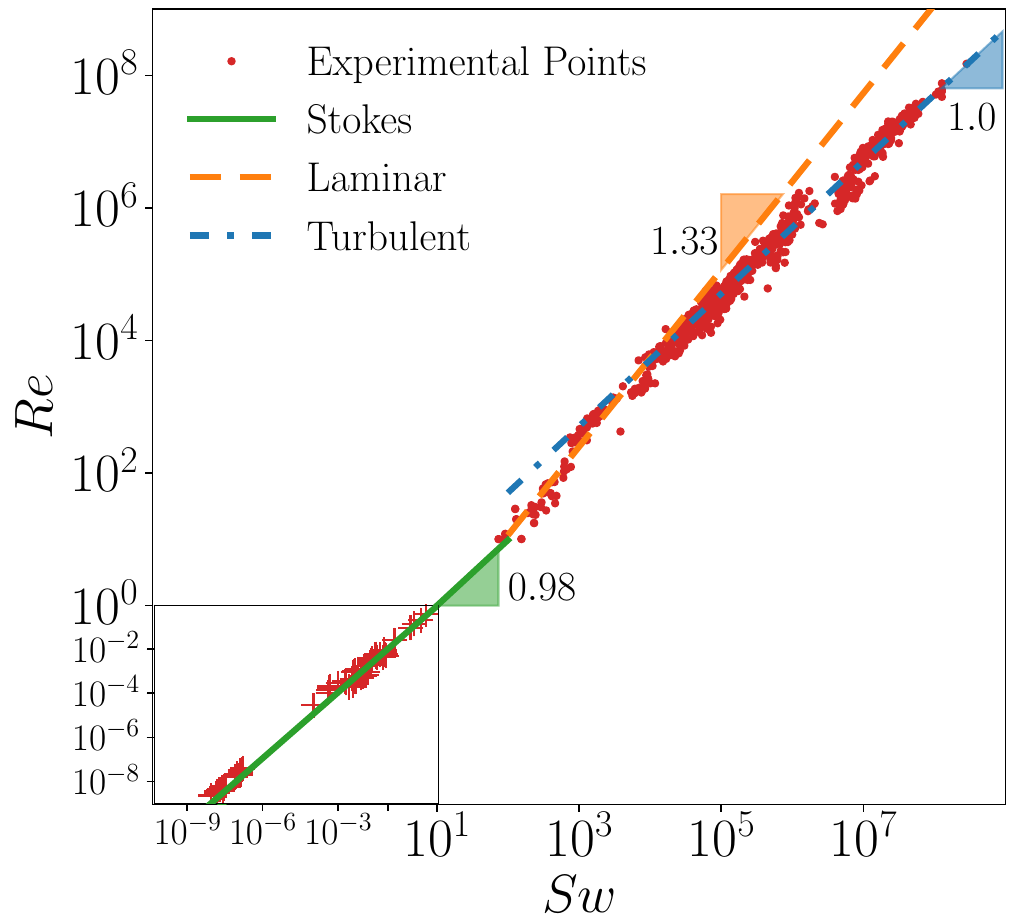}
    \caption{Reynolds number $Re$ as a function of the swimming number $Sw$. Crosses are new experimental data available in the database \cite{ventejou24DB}. Dots are experimental data collected by~\cite{Gazzola2014}. The three lines correspond to fitted curves and give the numerical scaling exponents of the three different regimes: $Re\sim  Sw^{0.98}$ for $Re<10$; $Re\sim Sw^{1.33}$ for $10<Re<10^4$; and $Re\sim Sw^{1.0}$ for $10^4<Re$.}
    \label{fig:RevsSw}
\end{figure}
The vortices created at the same frequency are spaced further apart as the Reynolds number (i.e. velocity) increases from $Re=100$ to $Re=960$. The length over which they dissipate becomes shorter towards the viscous regimes (Fig.~\ref{vorticityField}(c)), vanishing completely at low $Re$ (Fig.~\ref{vorticityField}(d)). Indeed, no vortices are present behind micro-organisms \cite{Drescher2010}.

%%\ppnote{Indeed, this could be included in a model where the social forces represent the cognitive target, namely turning a certain angle to avoid an obstacle or another swimmer. The curves in Fig.~\ref{fig:RvsThetaandF}, can be used as an \textit{abacus} to determine the force and angle required to turn the fish in a given direction.}

\textbf{Comparison with experimental data}. To compare our scaling results with experimental data, forces must be expressed in terms of observable data (Fig.~\ref{fig:RevsSw}), such as undulation or tail beating frequency. 
Although force $f_0$, torque $\tau_0$ and pulsation $\omega$ are independent quantities in the model, physical constraints exist between these quantities in living organisms.
We made the reasonable assumption that the size of the swimming organ scales with the size of the body $L$ \cite{Gazzola2014}. Force and torque generated by the swimming organ are such that $\tau_0 \sim f_0 L$. 

In inertial regimes, i.e. excluding the Stokes regime that we address separately, the instantaneous force creates a transient acceleration of the fluid, which scales with $L\omega^2$, \textit{i.e.} $f_0 \sim \rho L \omega^2$. 
Introducing the observation-based swimming number $Sw=\rho \omega L^2/\eta$~\cite{Gazzola2014}, the theoretical force drive can be related to experimentally measurable data as $Th\sim Sw^2$. 
The scaling laws previously derived from our model can thus be reformulated in terms of $Sw$: in the laminar regime $Re \sim Th^{2/3}\sim Sw^{4/3}$, while in the turbulent regime $Re \sim Th^{1/2} \sim Sw$, in accordance with the results of \citet{Gazzola2014}. 

\begin{figure*}[ht!]
    \centering
    \includegraphics[width=0.9\linewidth]{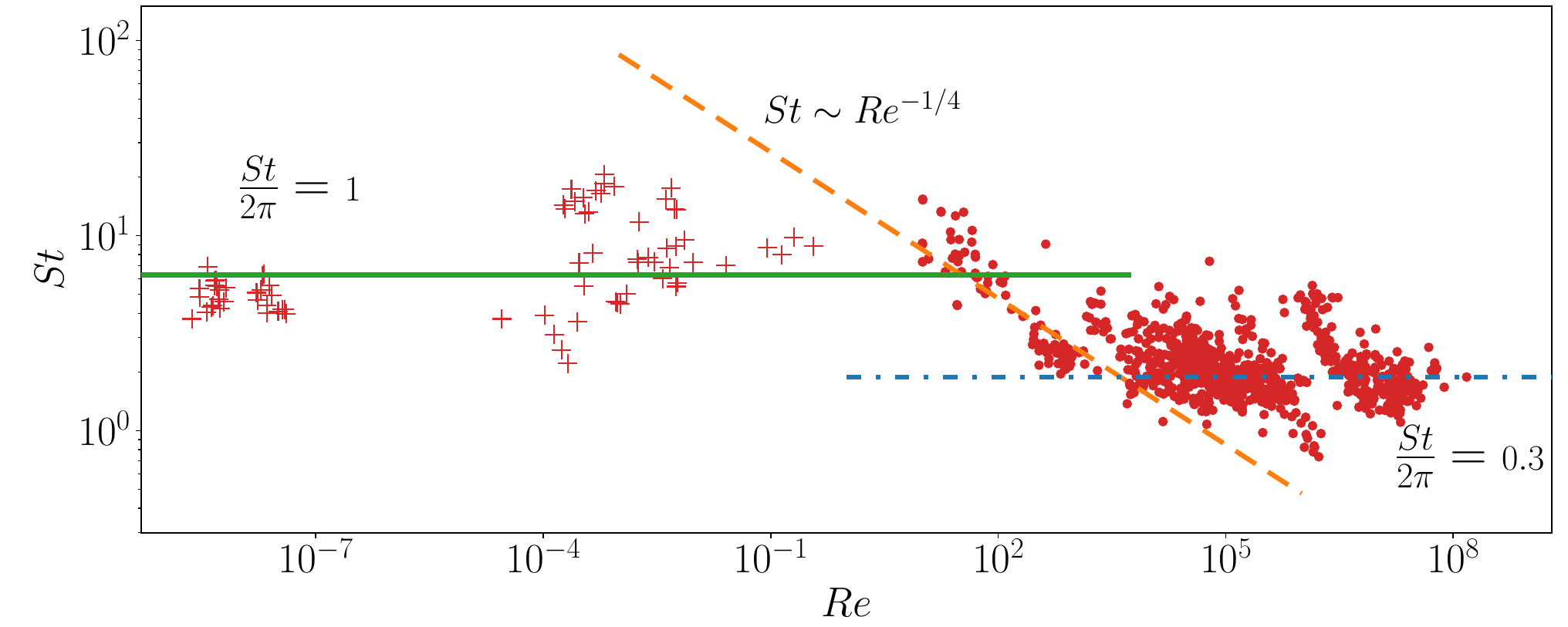}
    \caption{Strouhal number $St = A\omega / v$ as a function of the Reynolds number $Re$. Dots correspond to the data collected in~\cite{Gazzola2014}, and crosses are new experimental data. The three lines correspond to guides for the eyes with the three different regimes: $\tfrac{St}{2\pi}= 1$ for $Re<10$; $St\sim Re^{-0.25}$ for $10<Re<10^4$; and $\tfrac{St}{2\pi}\approx 0.3 $ for $10^4<Re$.}
    \label{fig:Stvsre}
\end{figure*}
In the Stokes regime, the force exerted by the swimming organ is balanced by viscous drag, so that $f_0 L^3 \sim \eta L v \sim \eta L^2 \omega$. 
Reintroducing again the swimming number $Sw$, we obtain $Th\sim Sw$, leading to $Re \sim Sw$.
Note that this is a natural consequence of the absence of inertia: each stroke creates a net displacement that scales with~$\sim L$, inducing a swimming velocity $v \sim L \omega$. 

Figure~\ref{fig:RevsSw} shows the experimental data collected from micrometer- to meter-size aquatic organisms
% \cite{bainbridge1958speed,hunter1971swimming,webb1986kinematics,webb1988steady,wardle1989muscle,videler1984fast,webb1982swimming,rosenberger1999functional,horner2008effects,fish1998comparative,fish1988kinematics,kojeszewski2007swimming,goldbogen2006kinematics,rohr2004strouhal,calambokidis1997blue,sato2010scaling,clark1979kinematics,wassersug1985kinematics,fish1984kinematics,muller2004swimming,green2011movement,budick2000locomotor,mchenry2005morphology,mchenry2003kinematics,brackenbury2004kinematics,woolley2001study,rikmenspoel1965tail,brokaw1966effects,brokaw1965non,brokaw1966effects,brokaw1965non,garcia2013hydrodynamique,sznitman2010,jordan1998,chattopadhyay2006swimming} 
along with the corresponding fitted scaling laws. In addition to the results from \citet{Gazzola2014}, an excellent agreement between experimental data and hydrodynamic scaling laws is also obtained in the Stokes regime.

Note that $v \sim L\omega$ in both the Stokes and turbulent regimes, so that the corresponding Strouhal number $St= Sw/Re$ is constant, as illustrated in Figure~\ref{fig:Stvsre}. The transition between  $St/2\pi\approx 0.3$~\cite{Gazzola2014} and $St/2\pi \approx 1$  occurs in the laminar regime where~$St \sim Re^{-1/4}$.

{\bf Conclusion}. By avoiding direct fluid-structure coupling, our generic swimmer model provides an efficient model for hydrodynamic propulsion, while retaining the salient features of swimming organisms across several decades of Reynolds numbers. High numerical stability and efficiency ensure fully tractable 2D and 3D simulations, thereby paving the way to large scale simulations with hundreds of agents. 
It also proposes a methodology for progressive refinement of the hydrodynamic field, by retaining higher moments of forces and torques, resulting in more complex propulsion models. This approach broadens our understanding of the swimming of aquatic organisms by revealing the universal relationship between the velocity of a swimmer and the force exerted by its swimming organ. 
The sub-linear dependence demonstrated between $Re$ and $Th$ suggests diminishing swimming performance as the swimmer’s Reynolds number increases. 
The scaling laws obtained also match the experimental data obtained from thousands of aquatic animals, ranging from large mammalians to micro-organisms. 
Our results shed new light on the general mechanisms underlying swimming and provide an efficient and robust numerical framework to investigate the collective behavior of swimmers in complex environments.

% \ppnote{It could be used as a building block to describe a school of fish. Indeed, in this case, the precise description of an individual fish's swim is less important as long as the fish are further apart than the size of their body undulation. }

% \ppnote{Our model also reproduces the characteristic wake of vortices \cite{Drucker2002} that is known to play a major role among fish in a shoal \cite{Ko2023}. And since our model swimmer also possesses steering properties and is able to perform controllable trajectories, it paves the way to large-scale analyses \adnote{(simulations?)} of swimming in complex environments, with for instance obstacles or other swimmers.}
% The model could help a lot to understand the emergent collective dynamics linked to both cognitive and physical (i.e. hydrodynamic) interactions.

{\bf Acknowledgements.} This project received financial support from the French National Research Agency (ANR-21-CE45-0005, FISHSIF project).

\bibliography{bibi}

\begin{thebibliography}{34}
\bibitem [{ven({\natexlab{b}})}]{ventejou24DB}%
  \BibitemOpen
  \href@noop {} {\enquote {\bibinfo {title} {Doi will be added after the peer
  review},}\ } ({\natexlab{b}})\BibitemShut {NoStop}%
\end{thebibliography}

\end{document}